\documentclass[12pt]{article}

\newread\testifexists
\def\GetIfExists #1 {\immediate\openin\testifexists=#1
    \ifeof\testifexists\immediate\closein\testifexists\else
    \immediate\closein\testifexists\input #1\fi}

\usepackage{gthstyle}
\immediate\openin\testifexists=e
    \ifeof\testifexists\immediate\closein\testifexists\else
    \immediate\closein\testifexists\input e\fipsf

\def\Bbb#1{\setbox0=\hbox{$\tt #1$}  \copy0\kern-\wd0\kern .1em\copy0}

\def\bbf#1{\setbox0=\hbox{$#1$} \kern-.025em\copy0\kern-\wd0
        \kern.05em\copy0\kern-\wd0 \kern-.025em\raise.0433em\box0}

\immediate\openin\testifexists=a
    \ifeof\testifexists\immediate\closein\testifexists\else
    \immediate\closein\testifexists\input a\fimssym.def          

       \def\b{\beta}         
        
         \def\k{\kappa}      
\def\m{\mu}                     \def\vv{\varphi}
         \def\j{\psi}    
\def\r{\varrho}     \def\s{\sigma}  
\def\t{\tau}        \def\th{\theta}  
              
      \def\W{\Omega}

 \def\ra{\rightarrow}

\def\dd{{\rm d}}  \def\bra{\langle}   \def\ket{\rangle}

\def\fract#1#2{{\textstyle{#1\over#2}}}
\def\ffract#1#2{\raise .3 em\hbox{$\scriptstyle#1$}\kern-.25em/
                \kern-.2em\lower .2 em \hbox{$\scriptstyle#2$}}

\def\part#1#2{{\partial#1\over\partial#2}}

\def\iz{\quad = \quad}

\newcommand{\be}{\begin{eqnarray}}
\newcommand{\ee}{\end{eqnarray}}
\newcommand{\eqn}[1]{(\ref{#1})}

\newcommand{\rf}[1]{\cite{ref:#1}}
\newcommand{\rr}[1]{\bibitem{ref:#1}}

\newcommand{\fn}{\footnote}

\newcommand{\newsec}[1]{\section{#1}\setcounter{equation}{0}}

   \newcommand{\el}[1]{\label{#1}\end{eqnarray}}

\begin{document}


\title{
\normalsize \hfill ITP-UU-04/01  \\ \hfill SPIN-04/01
\\ \hfill {\tt gr-qc/0401027}\\ \vskip 20mm
\Large\bf
Horizons
\author{Gerard 't~Hooft}
\date{\small Institute for Theoretical Physics \\
Utrecht University, Leuvenlaan 4\\ 3584 CC Utrecht, the
Netherlands\medskip \\ and
\medskip \\ Spinoza Institute \\ Postbox 80.195 \\ 3508 TD
Utrecht, the Netherlands \smallskip \\ e-mail: \tt
g.thooft@phys.uu.nl \\ internet: \tt
http://www.phys.uu.nl/\~{}thooft/}}

\maketitle

\begin{quotation} \noindent {\large\bf Abstract } \medskip \\
The gravitational force harbours a fundamental instability against
collapse. In standard General Relativity without Quantum
Mechanics, this implies the existence of black holes as natural,
stable solutions of Einstein's equations. If one attempts to
quantize the gravitational force, one should also consider the
question how Quantum Mechanics affects the behaviour of black
holes. In this lecture, we concentrate on the horizon. One would
have expected that its properties could be derived from general
coordinate transformations out of a vacuum state. In contrast, it
appears that much new physics is needed. Much of that is still
poorly understood, but one may speculate on the way information is
organized at a horizon, and how refined versions of Quantum Theory
may lead to answers.

\end{quotation}
\bigskip\bigskip



\newsec{Introduction: Black Holes as Inevitable Features of
General\\ Relativity}
The fact that the gravitational force acts directly upon the
inertial mass of an object, makes this force unique in Nature,
and allows for an unambiguous description of the classical
(\emph{i.e.} unquantized) case, called ``General Relativity".
However, unlike the situation in electromagnetism, the
gravitational force produces attraction rather than repulsion
between like charges. An inevitable consequence of this is a
fundamental instability: masses attract to form bigger masses,
which attract one another even more strongly. Eventually,
gigantic implosions of large accumulated quantities of mass may
result. There is no obvious limit here, so one cannot avoid that
the gravitational potential might cross an important threshold,
where the escape velocity exceeds that of light.

Indeed, as soon as one is ready to accept the validity of General
Relativity for classical systems, one can easily calculate what
will happen. The final state that one then reaches is called a
"black hole". In astronomy, the formation of a black hole out of
one or several stars depends on the circumstances, among which is
the equation of state of the material that the stars are made of.
Because of this, the physics of black hole formation is sometimes
challenged, and conjectures are uttered that black holes "are
probably nothing else but commercially viable figments of the
imagination"\rf{MV}

It is however easy to see that such a position is untenable. To
demonstrate this, let me here show how to construct a black hole
out of ordinary objects, obeying non-exotic equations of state.
These objects could, for example, be television sets, acting on
batteries. During the process of black hole formation, these
objects will each continue to be in perfect working order. We
begin with placing these in the following configuration: let them
form a shell of matter, of thickness \(d\) and radius \(R\). If
\(d\) is kept modest, say a few kilometers, then \(R\) has to be
taken vary large, say a million light years. The television sets
may stay at convenient distances away from each other, say a
meter. The initial velocities are taken to be small; certainly
objects close to each other must have very small relative
velocities so that collisions have no harmful effects.

The objects attract one another. They don't feel it because,
locally, they are under weightless conditions, but they do start
to accelerate. So, the sphere shrinks. After thousands of years,
the configuration is still spherical, the relative velocities for
close-by objects are still small, the density is still low, the
televisions are still in working order, but they pass the magical
surface called ``horizon". What happens is, that light emitted by
the objects can no longer reach the outside world. The
calculation is straightforward and robust, which means that small
perturbations will not affect the main result: no light can be
seen from these objects; they form a black hole.

What happens next, is that the sphere of objects continue to
contract, and at some point, long after the horizon has been past,
the objects crush, television sets will cease to function, for a
while the Standard model still applies to them, but eventually the
matter density will exceed all bounds, and a true singularity is
encountered\fn{Small perturbations from spherical symmetry do
affect the singularity in a complicated way, but this is not
relevant for the nature of the horizon.}. It is here, at this
singularity, where the laws of physics as we know them no longer
apply, but whatever happens there is totally irrelevant for the
phenomenology of a black hole; whatever an outside observer sees
is determined by known laws of physics. the horizon acts as a
''cosmic sensor", preventing us from observing the singularity.
Whether all singularities in all solutions to the equations are
always screened by this cosmic sensor is still being debated, but
we do see this happen in all practical solutions known.

\begin{figure} \setcounter{figure}{0}
\begin{center} \epsfxsize=100 mm\epsfbox{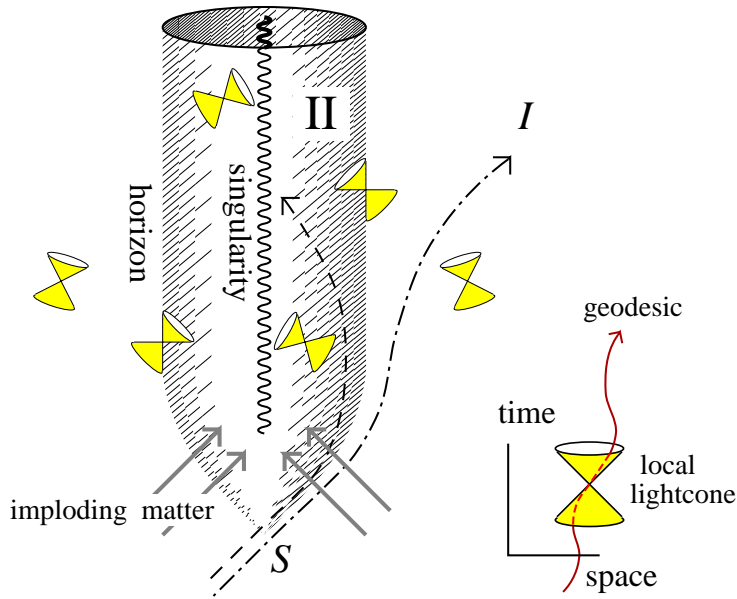}
  \caption{\small{The space-time of a black hole}\label{bhspacetime.fig}}
\end{center}
\end{figure}

In Fig.~\ref{bhspacetime.fig}, it is sketched what happens to
space-time. The solution to Einstein's equations in General
Relativity dictates that, locally, the light cones are tilted. The
shaded surface, the horizon, is formed by constructing the tangent
of these local light cones. Inside this surface, all local
lightcones are pointed inwards, towards the central singularity.
The radius of this horizon is found to be \be R=2G_NM/c^2\
,\el{radius} where \(M\) is the total mass-energy of the system.

Note that all signals seen by an outside observer in region \(I\),
when observing a black hole, originate at the point \(S\). If only
a finite amount of light is emitted from there, this light is
spread over an infinite amount of time, and therefore infinitely
red-shifted. Hence, one expects no signal at all; the black hole
is black. It can be formed out of ordinary matter.

\newsec{Black holes in particle physics}
In elementary particle physics, the gravitational force is
extremely weak, and can normally be ignored. It is, however,
understood that there must be regions of particle physics where
this force \emph{must} play a decisive role. this is when the
energy per particle tends to exceed the Planck scale. The Planck
scale is set by considering the three fundamental constants of
nature: \be \mathrm{The\ velocity\ of\ light,}\qquad\qquad\quad\
c&=&2.9979\times 10^{8}\ \ \mathrm{m/sec\ ,}\qquad\qquad\cr
\mathrm{Planck's\ constant,}\qquad\quad
h/2\pi=\hbar&=&1.0546\times 10^{-34}\ \mathrm{ kg\ m^2/sec\
 },\qquad\qquad \mathrm{and}\cr \mathrm{Newton's\
constant,}\qquad\qquad\quad\  G_N&=& 6.672\times 10^{-11}\
\mathrm{ m^3\ kg^{-1}\,sec^{-2}\ . }\quad\qquad\el{constants}
 Out of these, one finds the
following fundamental units: \def\pl{\mathrm{Planck}} \be L^\pl
\iz \sqrt{\hbar\,G_N/c^3}&=&1.616\times 10^{-33}\ \mathrm{cm}\
,\cr T^\pl\iz\sqrt{\hbar\,G_N/c^5}&=&5.39\times 10^{-44}\
\mathrm{sec}\ , \cr M^\pl\iz  \sqrt{\hbar\,c/G_N}&=&21.8\ \m
\mathrm{g}\ .\el{planck}

If particles collide with the enormous c.o.m. energy of \(M^\pl
c^2\), gravitational effects must be important. If many particles
with such energies accumulate in a region of size \(L^\pl\),
gravitational implosion must take place. Black holes must play an
important role there.

It was S.~Hawking's fundamental discovery \rf{SH}, that, when
applying the laws of Quantum Field Theory (QFT), black holes are
no longer truly black. Particles are emitted at a fundamental
temperature, given by \def\bh{\mathrm{BH}}\be
k\,T^\mathrm{Hawking}&=& {\hbar\ c^3\over
8\pi\,G_N\,M_\bh}\iz{\hbar\ c\over 4\pi R_\bh}\ .\el{Thawking}
For astronomical black holes, this temperature is far too low to
give observable effects, but in particle physics the Hawking
effect is essential. For a further discussion of this phenomenon
by the present author, see \rf{Sm}.

One might suspect now that black holes therefore behave a bit
more like ordinary matter. Not only can they be formed by
high-energy collisions, but they can also decay. Apparently, QFT
restores (some) time-reversal symmetry in processes involving
black holes. Are black holes elementary particles? Are elementary
particles black holes? Probably, particles and black holes become
indistinguishable at the Planck scale. It is instructive to
consider the entire formation and decay of a black hole as if
described by quantum mechanical amplitudes.
\newsec{Information in a black hole}
\def\inn{\mathrm{in}}\def\out{\mathrm{out}}
The absorption cross section \(\s\) is roughly given by
\be\s=2\pi R_\bh^2=8\pi M_\bh^2\ ,\el{sigma} and the emission
probability \emph{for a single particle in a given quantum state}:
\be W\dd t={\s(\mathbf{k})\,v\over V}\,e^{-E/k\,T}\ \dd t\
 ,\el{emission} where \(\mathbf k\) is the wave number
characterizing the quantum state of the particle emitted, and
\(T\) is the Hawking temperature. \(E\) is the energy of the
emitted particle. Now, \emph{assume} that the process is also
governed by a Schr\"odinger equation. this means that there are
quantum mechanical transition amplitudes, \be { \cal T}_\inn&=&{
}_\bh\bra M+E/c^2|\ |M\ket_\bh|E\ket_\inn\ ,\label{inampl} \\
{\cal T}_\out&=&{ }_\bh\bra M|_\out\bra E|\ |M+E/c^2\ket_\bh\
,\el{outampl} where \(|M\ket_\bh\) is the black hole state
without the absorbed particle, having mass \(M\), and
\(|M+E/c^2\ket\) is the slightly heavier black hole with the
extra particle absorbed. The absorption cross section is then \be
\s=|{\cal T}_\inn|^2\r(M+E/c^2)/v\ ,\el{qmsigma} where
\(\r(M+E/c^2)\) is the level density of the black hole in the
final state. This is what we get when applying Fermi's Golden
Rule. The same Golden Rule gives us for the emission process
\emph{for each quantum state of the emitted particle}: \be
W=|{\cal T}_\out|^2\r(M){1\over V}\ .\el{qmemission} Here, as
before, \(v\) is the velocity of the emitted particle, and \(V\)
is the volume, to be included as a consequence of the
normalization of the quantum state.

We can now divide Eq.~\eqn{sigma} by Eq.~\eqn{emission}, and
compare that with what we get when \eqn{qmsigma} is divided by
\eqn{qmemission}. One finds: \be
{\r(M+E/c^2)\over\r(M)}=e^{E/kT}=e^{8\pi G_N M\,E/\hbar\,c^3}\ .
\el{divide} One concludes that \be \r(M)&=&e^{S(M)}\ ,\cr S(M+\dd
M)-S(M)&=& 8\pi G_N M \dd M/\hbar\,c \ ;\\
S(M)&=&{4\pi\, G_N\over \hbar\,c} M^2 +C^\mathrm{\,nt}\ .\el{S}
Thus, apart from an overall multiplicative constant,
\(e^{C^\mathrm{\,nt}}\), we find the \emph{density of states}
\(\r(M)=e^{S(M)}\) for a black hole with mass \(M\). It can also
be written as \be \r(M)=2^{A/A_0}\ ,\el{bits} where \(A\) is the
area \(4\pi R^2\) of the black hole, and \(A_0\) is a fundamental
unit of area, \be A_0=0.724\times 10^{-65}\ \mathrm{cm}^2\
.\el{bitarea} Apparently, the states of a black hole are counted
by the number of bits one can put on its horizon, one bit on every
\(A_0\).

This result is quite general. It also holds for black holes that
carry electric charge or angular momentum or both. Usually, one
expects the constant \(C^\mathrm{\,nt}\) in Eq.~\eqn{S} to be
small, although its value is not known.
\newsec{The Brick Wall}
This result\rf{entropy}, obtained in the 1970's, is astounding.
Black holes come in a denumerable set of states. These states
seem to be situated on the horizon, and, as was stated in the
Introduction, the physical properties of the horizon follow from
simple coordinate transformation rules applied on the physical
vacuum. We seem to have hit upon a novel property of the vacuum
itself.

Naturally, we wish to learn more about these quantum states. It
should be possible now to derive all their properties from
General Relativity combined with Quantum Field Theory. However,
when one tries to do these calculations, a deep and fundamental
mystery emerges: direct application of QFT leads to an infinity
of states, described by much more parameters than one bit of
information per quantity \(A_0\) of area. Writing the radial
coordinate \(r\) and the external time coordinate \(t\) as \be
r=2M+e^{2\s}\ ;\qquad t=4M\t\ ,\el{expcoord} in units where all
Planckian quantities of Eq.~\eqn{planck} were put equal to one,
it is quickly found that, at the horizon, in-going and out-going
waves are plane waves in terms of \(\s\) and \(\t\):
\be\j(\s,\t)\ra \j_\inn(\s+\t,\,\W)+\j_\out(\s-\t,\,\W)\ ,
\el{planewaves} where \(\W\) stands short for the angular
coordinates \(\th\) and \(\vv\) on the horizon. Since \(\s\) runs
to \(-\infty\), an infinite amount of information can be stored
in these waves.

By way of exercise, one can now compute how much information will
be stored in these waves if \begin{itemize}\item[-] the particle
contents will be as dictated by the Boltzmann distribution
corresponding to the Hawking temperature \eqn{Thawking}, and
\item[-] a \emph{brick wall} is placed at some postion \(r_w=2M+h\),
where some boundary condition is imposed on the fields. One
could impose a Neumann or Dirichlet boundary condition for the
fields there, or something more sophisticated\fn{Since one
expects \emph{all} continuous symmetries to be broken by the
black hole, a \emph{random} boundary condition could be
preferred, but in practice the details of the boundary condition
are not very important.}.
\end{itemize}
In a theory with \(N\) scalar fields, in the limit of small \(h\)
one finds\rf{Sm,ref:GtH85} for the total energy of the particles: 
\be U= {2\pi^3\over 15 h}\left({2M\over\b}\right)^4N\ ,
\el{energywall} and for the total entropy: \be S={16\pi^3M\over 45
h}\left({2M\over\b}\right)^3N\ .\el{entropywall} We can place the
wall in such a way that the entropy matches Eq.~\eqn{S}:\be
h={N\over 720\pi M}\ .\el{wallposition} The total energy of the
particles then makes up for \(\fract38\) of the black hole mass.

Only with this brick wall in place, a black hole would exactly
live up to our intuitive expectations. Infalling waves would
bounce back, so that an unambiguous \(S\)-matrix can be derived,
and the entropy \(S\) would correspond to the total number of
physical states. Although the wall position may seem to depend on
the total mass \(M\) of the black hole, one finds that the
\emph{covariant} distance between wall and horizon is \(M\)
independent: \be\int_{r-2M}^{r=2M+h}\dd s=\sqrt{N\over 90\pi}\ .
\el{invarwall}

But what would be the physical interpretation of this result?
Surely, an infalling observer would not notice the presence of
such a wall. For some reason, a quantum field cannot have
physical degrees of freedom between the wall and the horizon, but
why?

One obvious observation is that this is a region comparable to the
Planck size (or even somewhat smaller). Surely, one is not allowed
to ignore the intense gravitational self interactions of particles
confined to such a small region, so that perturbative quantum
field theory probably does not apply there. However, one could
concentrate \emph{only} on either the in-going or the out-going
particles. They are just Lorentz transforms of regular states. Why
should their degrees of freedom no longer count?

A more subtle suggestion is that, although we do have fields
between the wall and the horizon, which do carry degrees of
freedom, these degrees of freedom are not physical. They could
emerge as a kind of \emph{local gauge degrees of freedom},
undetectable by any observer. Such a suggestion ties in with what
will be discussed later (Section~\ref{infoloss.sec}).
\newsec{The black hole caustic}
\begin{figure}
\begin{center} \epsfxsize=85 mm\epsfbox{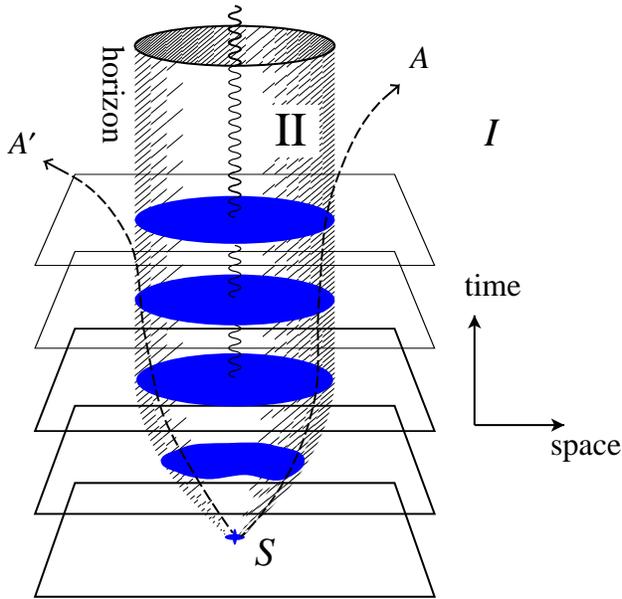}
  \caption{\small{The black hole caustic}\label{bhcaustic.fig}}
\end{center}
\end{figure}
One can do a bit more than speculate. In Ref.~\rf{Sm}, it is
described how to take into account that in-going and out-going
particles interact gravitationally. We know now that such
interactions may not be ignored. What is found is that the
position of the horizon depends on the mass distribution of
matter falling in. In turn, this affects the quantum states of
the particles moving out, so, unlike what one finds from
linearized quantum field theory, there is a relation between
in-going and out-going particles, and this relation can indeed be
cast in the form of an \(S\)-matrix. The problem with this
derivation is that one still does not find the correct density of
distinct quantum states --- there are too many states. It turns
out that the quantum state of out-going matter appears to be
described completely by the \emph{geometry of the dynamic
horizon}.

This can be understood in the following way. We define the
horizon as the boundary between the region \(I\) of space-time
from where signals can escape to infinity, and the region \(II\)
from which no signals can be received. This means that the exact
location of the horizon is obtained by following all light rays
all the way to time \(\ra +\infty\). if we have obtained the
horizon at some (large) value of time \(t\), we can integrate the
equations to locate the horizon at earlier times. The procedure
is sketched in Fig.~\ref{bhcaustic.fig}. If, finally, we reach
the instant when the black hole is formed, the horizon shrinks,
until the region \(II\) ends. The horizon opens up at the point
\(S\), but, from its definition, we see that, there, it is a
caustic. The details of this caustic can be quite complex, in
principle. Indeed, the quantum state of the out going particles
depends on where on this caustic these particles originated. One
might conclude that the quantum state of the out-going particles
might be determined completely by the geometric features of this
caustic.

As yet, however, it has not been possible to distill a Hilbert
space of out-going particles from such an assumption. In
Fig.~\ref{bhcaustic.fig}, we see that a signal observed by
observer \(A\), may meet the signal seen by an observer \(A'\) at
the caustic. \(A\) and \(A'\) need not be opposite to one
another; there is a duality mapping from points \(A\) to points
\(A'\) on the horizon. This mapping may be the one that
determines the black hole's quantum state.

A particle seen by \(A\) and a particle seen by \(A'\) meet at
\(S\) with tremendous c.o.m. energy. Let us consider their
scattering in a time-reversed setting. Gravitational interactions
cause both particles to undergo a large coordinate shift\rf{DH}.
These shifts turn both particles into showers (backward in time).
The quantum states of \(A\) and \(A'\) are determined by
overlapping these showers with the vacuum fluctuations in the
region below \(S\).
\newsec{Strings from black holes; white holes}
How gravitational interactions lead to string-like amplitudes for
the entire process of black hole formation and evaporation, has
been described in refs.~\rf{Sm} and \rf{BhStr}. It is not exactly
string theory what one gets, but rather a string with a purely
imaginary string constant. Since the horizon itself acts as the
string world sheet, this string may be some dual of the
conventional string approach to black holes\rf{P}. One can
picture the scattering events as follows. The black hole is
formed by a large number of particles contracting. Each of these
particles is pictured as a closed string. Since the horizon acts
as the string world sheet, our closed strings widen as they
approach the horizon, and they scan the entire horizon as they do
so. The strings recombine to form new closed strings, which then
separate from the horizon acting as Hawking particles. A regular
space-time with an expanding cloud of tiny closed strings forms
the final state.

A peculiar by-product of this analysis is the resolution of an
old problem in black hole physics: what is the time-reverse of a
black hole? In the literature it is sometimes known as the
``white hole": a shrinking black hole emitting classical objects
and eventually disappearing. It may have been formed by a cloud
of ``time-reversed Hawking particles".

In our analysis the answer is as follows. By assuming that the
out-state is controlled by the in-state through gravitational
interactions, it is found that the amplitude automatically
respects time-reversal invariance, basically because the
gravitational forces respect Newton's law: action \(=\) reaction.
It is found that the \emph{positions} of the out-going particles
are determined by the \emph{momenta} of the in-going ones, and
\emph{vice-versa}. Quantum mechanically, the particles in the
momentum representation are superpositions of the particles in the
position representation. Therefore, one finds that \emph{white
holes are quantum superpositions of all possible black hole
states} (in the same mass region), \emph{and vice-versa}.
\newsec{Information loss\label{infoloss.sec}}
Much of the investigations described above pertains to an
apparent incongruity in any quantum picture of black holes.
Classically, one sees that objects falling in cannot imprint all
information contained in them on the out-going states. They are
described by quantum waves that require an infinite amount of
time to enter the black hole. In contrast, the out-going
particles were there already at the very beginning, waiting close
to the horizon, at \(\s\) in the far negative region, until it is
their turn to leave. Our quantum picture requires that these
objects are nevertheless closely related. The analysis sketched
in the previous sections might suggest that we have come close to
resolving this problem: all one has to do is switch on the
gravitational forces between in-going and out-going objects.
String theory\rf{P} also suggests that this problem can be cured.

However, it should be possible to identify these quantum states in
terms of features of the vacuum in relation to general coordinate
transformations. In particular, this should be possible for the
horizon in the large mass limit. The space-time one then
describes is known as Rindler space-time\rf{R}. What seems to be
missing is the identification of the quantum states in Rindler
space-time and their relation to the quantum states
characterizing the vacuum in a flat world. This flat world
appears to allow for an unlimited amount of information to
disappear across the horizon. To see this, all one has to do is
subject ordinary particles to unlimited Lorentz boost
transformations. In spite of all that has been said, this problem
has not been solved in a satisfactory manner.

Since we are dealing here with quantum phenomena in an extremely
alien world of highly curved coordinate frames, it is natural to
ask the following question: \emph{Why should these issues
\emph{not} be related to the question of the foundation of quantum
mechanics?} There are more than just instinctive reasons to ask
this question. As soon as one allows space and time to be curved,
one has to confront the possibility that they form a closed,
finite universe. Of course, quantum cosmology must be a legitimate
domain of quantum gravity. But the formulation of the quantum
axioms for closed universes leads to new difficulties. One of
these is the fact that there is no external time coordinate, which
means that one will not have transition amplitudes or
\(S\)-matrices. One then encounters the difficulty of
interpretation: if the universe is finite, one cannot repeat an
experiment infinitely many times at far separated places, so, if a
quantum calculation leads to the evaluation of a ``probability",
how then can one verify this? In this universe, something happens
or it does not, but probabilistic predictions then amount to
imprecise predictions. Must we accept an imprecise theory? This
difficulty shows up quite sharply in simple ``model universes",
such as the one formed by gravitating particles in 2 space-, 1
time dimension. This is a beautiful model with only a finite
number of physical degrees of freedom\rf{twodimgrav}, so
quantization should be straightforward; unfortunately, it is not,
and the fore-mentioned difficulties are the reason.

Should we return to the old attempts at constructing ``hidden
variable theories" for quantum mechanics?\rf{B} Usually, such
endeavor is greeted with skepticism, for very good reasons. Under
quite general assumptions, it has been demonstrated that:
``hidden variables cannot be reconciled with locality and
causality".

This would indeed be a good reason to abandon such attempts. But,
how general is this result? In Ref.~\rf{determ}, some very simple
models are constructed that could be viewed as counter examples
of the general theorem. We hasten to add that these model are not
at all free from problems. One might suspect, however, that the
well-known no-go theorems for hidden variables do rely on some
assumptions, which seem to be so natural that one tends to forget
about them. Here, we list some of the small-print that may have
gone into the derivation of the argument: \begin{itemize}
\item[-] It was assumed that an observer at all times is free to
choose from a set of non-commuting operators, which of these
(s)he wishes to measure. \item[-] Rotations and other continuous
symmetry operations can be performed locally, without disturbing
any of the quantum states elsewhere. \item[-] The vacuum is a
single, unique state. \end{itemize} Assumptions of this kind may
actually not be valid at the Planck scale. Indeed, in
Ref.~\rf{determ} it is assumed that only one class of operators
can truly be observed at the Planck scale, and they all commute.
They were called `beables' there.

The most important problem of the ones just alluded to is that
deterministic evolution seems to be difficult to reconcile with a
Hamiltonian that is \emph{bounded from below}.  It is absolutely
essential for Quantum mechanics to have a lowest energy state,
\emph{i.e.}, a vacuum state. Now the most likely way this problem
can perhaps be addressed is to assume not only deterministic
evolution, but also \emph{local information loss}. As stated,
information loss is difficult to avoid in black holes, in
particular when they are classical. it now seems that this indeed
may turn up to be an essential ingredient for understanding the
quantum nature of this world.

Simple examples of universes with information loss can be modeled
on a computer as cellular automata\rf{W}. An example is `Conway's
game of life'\rf{C}.

Information loss may indeed already play a role in the Standard
Model! Here, \emph{local gauge degrees of freedom} are pieces of
information that do play a role in formulating the dynamical
rules, but they are physically unobservable. An unorthodox
interpretation of this situation is that these degrees of freedom
are unobservable \emph{because} their information contents get
lost, much like information drained by a black hole.

We stated earlier that the fields between the horizon and the
brick wall could be local gauge degrees of freedom. Now we can
add to that that probably they represent lost information.
\newsec{Freezing the horizon}
String theory has produced some intriguing insights in the nature
of the black hole microstates. Unfortunately, the results
reported apply to exotic versions of black holes, and the
ordinary Schwarzschild black hole is conspicuously absent. Yet it
is the Schwarzschild black hole that we hold here as the
prototype. What went wrong?

The black holes handled in string theory are all \emph{extreme
black holes} or close-to-extreme black holes. What is an extreme
black hole?

The prototype of that can be obtained from the Reissner-Nordstr\o
m black hole, a black hole with a residual electric charge. Due
to the stress-energy tensor of the electric field, the metric is
modified into \be \dd s^2=-\left(1-{2M\over r}+{Q^2\over
r^2}\right)\dd t^2+{\dd r^2\over 1-{2M\over r}+{Q^2\over
r^r}}+r^2\dd\W^2\ ,\el{RN} where \(M\) is the mass, as before, and
now \(Q\) is the electric charge (in Planck units). As long as
\(Q<M\), the quantity \(\,1-{2M\over r}+{Q^2\over r^2}\ \) has two
zeros, \be r_\pm=M\pm\sqrt{M^2-Q^2}\ .\el{zeros} The largest of
these is the location of the horizon. The smaller value
represents a second horizon, which is hidden behind the first.
The physical interpretation of these horizons is nicely exhibited
in Ref.~\rf{HE}, but does not concern us here very much. The
\emph{extreme case} is when \(Q\) approaches \(M\). Then, the
quadratic expression becomes \((1-{M\over r})^2\), and it has two
coinciding horizons at \(r=r_+=r_-=M\). It is this kind of
horizon that can be addressed by string theories.

Actually, the situation is a little bit deceptive. One could
argue that, in the extreme limit (which can probably be
approached physically, but never be reached exactly), the two
horizons do not coincide at all. This is because, if we follow a
\(t=\)constant path from \(r_+\) to \(r_-\), the metric distance
between the horizons becomes \be
\int_{r_-}^{r^+}\sqrt{-g_{11}}\dd r\ra M\pi\ ,\el{hordistance}
and this does not tend to zero in the limit (and it is a time-like
distance, not space-like). Moreover, the distance between the
\(r_+\) horizon and any point in the regular region \(I\) of the
surrounding universe (say the point \(r=2M\)), tends to infinity:
\be\int_{r_+}^{2M}\sqrt{g_{11}}\ra \infty\ .\el{horinfty}

In the extreme limit, the horizon is also infinitely red-shifted,
the gravitational field \(\k\) there tends to zero, and so does
the Hawking temperature. In all respects, the extreme horizon is
a \emph{frozen horizon}. Its surface area is still \(4\pi r_-^2 =
4\pi M^2\ne 0\). Accordingly, the entropy \(S=\pi M^2\ne 0\).
However, sometimes it is argued that extreme black holes should
have vanishing entropy. What happened?

\begin{figure}
\begin{center} \epsfxsize=80 mm\epsfbox{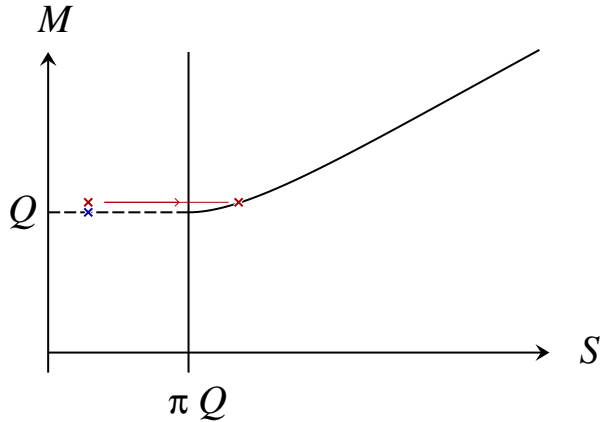}
  \caption{\small{The energy plotted against entropy at fixed \(Q\).
  The dotted horizontal line does not follow from Eq.~\eqn{RNenergy},
  but from physical considerations.}\label{rncurve.fig}}
\end{center}
\end{figure}

The entropy for the Reissner-Nordstr\o m black hole is \be S=\pi
r_+^2=\pi\Big(M+\sqrt{M^2+Q^2}\Big)^2\ .\el{RNentropy} Inverting
this gives the mass-energy \(M\) as a function of \(Q\) and
\(S\): \be M={Q\over 2}\left(\sqrt{\pi Q\over
S\vphantom{Q}}+\sqrt{ S\over \pi Q}\,\bigg)\quad,\quad
T={\partial M\over\partial S}\right|_Q \ .\el{RNenergy} This
curve, at a fixed value for \(Q\), is sketched in
Fig.~\ref{rncurve.fig} It begins at \(S=\pi Q\) since, as we see
from \eqn{RNentropy}, \(S\geq\pi Q\). At the extreme point, the
temperature \(T\) is zero.

Our physical intuition, however, tells us that perhaps states
with more order in them also exist, so that one can indeed lower
\(S\). The temperature will not become less than zero, so one
simply expects a straight horizontal line from \(S=0\) to \(S=\pi
Q\) (dotted line in Fig.~\ref{rncurve.fig}). One might suspect
that, in a superior quantum theory, tiny deviations from the
straight line may occur.

Now what happens if we take one of these ordered states, with
\(S\ll \pi Q\) (lowest cross in Fig.~\ref{rncurve.fig}), and cause
a minor disturbance, for instance by throwing in a light neutrino?
The energy rises slightly (cross top left), and the hole will no
longer be extreme . However, the correct solution is now the
position on the curve at the right. Complete disorder must take
place (arrow). Apparently, the slight perturbation from the
neutrino now caused complete disorder. This can be understood in
simple models. Since the horizon is no longer extreme, it is also
no longer frozen. Dynamical evolution sets in, and this causes
disorder. The situation can again be modeled in simple cellular
automata.
\newsec{Conclusion}
With some physical intuition, one can view the horizon of a black
hole as an intriguing physical object. Its microstates as yet
cannot be linked to local properties of the vacuum configuration
out of which the horizon is transformed, but string theory has
made progress in picturing frozen or slowly evolving horizons. In
principle, what has been discussed here should also apply to
horizons in different settings, such as cosmological horizons.
Considerable caution is then asked for, however, since quantum
mechanics might not apply to an entire cosmos.

\end{document}